\documentclass{aa}  
\usepackage{graphicx,natbib}
\usepackage{txfonts}
\def\kms{km~s$^{-1}$}

\newcommand{\nc}{\newcommand}

\nc{\RAJ}[4]{$\alpha(J2000) = {#1}^{\rm h}{#2}^{\rm m}{#3}\fs{#4}$}
\nc{\DecJ}[4]{$\delta(J2000) = {#1}\degr {#2}\arcmin {#3}\farcs{#4}$}

\begin{document}

   \title{Maser emission from the CO envelope of the asymptotic giant branch star W Hydrae}

  \titlerunning{CO maser emission around W Hya}

%   \subtitle{}

   \author{W.~H.~T. Vlemmings
          \inst{1}\fnmsep\thanks{wouter.vlemmings@chalmers.se}
          \and
          T. Khouri\inst{1}
                    \and
          D. Tafoya\inst{1}
          }

   \institute{Department of Space, Earth and Environment, Chalmers University of Technology, Onsala Space Observatory, 439 92 Onsala, Sweden
}

   \date{09-Sept-2021}

% \abstract{}{}{}{}{} 
% 5 {} token are mandatory
 
  \abstract
  % context heading (optional)
  % {} leave it empty if necessary  
   {Observation of CO emission around asymptotic
     giant branch (AGB) stars is the primary method to determine gas mass-loss rates. While radiative transfer models have shown that molecular levels of CO can become mildly inverted, causing maser emission, CO maser emission has yet to be confirmed observationally.}
  % aims heading (mandatory)
   {High-resolution observations of the CO emission around AGB
     stars now have the brightness temperature sensitivity to detect possible weak CO maser emission.}
  % methods heading (mandatory)
   {We used high angular resolution observations taken with the Atacama Large Millimeter/submillimeter Array (ALMA) to observe the  small-scale structure of CO~$J=3-2$ emission around the oxygen-rich AGB star W~Hya.}
  % results heading (mandatory)
   {We find CO maser emission amplifying the stellar continuum with an optical depth $\tau\approx-0.55$. The maser predominantly amplifies the limb of the star because CO~$J=3-2$ absorption from the extended stellar atmosphere is strongest  towards the centre of the star. }
  % conclusions heading (optional), leave it empty if necessary
   {The CO maser velocity corresponds to a previously observed
     variable component of high-frequency H$_2$O masers
     and with the OH maser that was identified as the amplified
     stellar image. This implies that the maser originates beyond
     the acceleration region and constrains the velocity profile since we find the population inversion primarily in the inner circumstellar envelope. We find that inversion can be explained by the radiation field at 4.6~$\mu$m and that the existence of CO maser emission is consistent with the estimated mass-loss rates for W~Hya. However, the pumping mechanism requires a complex interplay between absorption and emission lines in the extended atmosphere. Excess from dust in the circumstellar envelope of W~Hya is not sufficient to contribute significantly to the required radiation field at 4.6~$\mu$m. The interplay between molecular lines that cause the pumping can be constrained by future multi-level CO observations.}


   \maketitle
%
%-------------------------------------------------------------------

\section{Introduction}
The mass loss of asymptotic giant branch (AGB) stars, with
main-sequence masses in the range $1-8$~M$_\odot$, is an important source
of interstellar enrichment \citep{HO18}. The primary method of
determining the gas mass-loss rates of these stars is to use observations of
CO lines in their circumstellar envelopes (CSEs). Radiative transfer
calculations of rotational and vibrational CO transitions in the
(sub)millimetre to far-infrared (far-IR) wavelength range are subsequently
used to determine the total gas content in the CSE, and thus,
indirectly, the AGB mass-loss rate. 

It has long been known that, under specific conditions in CSEs, some
of the lowest rotational energy levels of CO can become mildly
inverted causing weak maser emission
\citep[e.g.][]{Goldsmith72,Morris80,Schoenberg88, Piehler91}. Masing
of CO emission could also occur in shocks in molecular clouds
\citep[e.g.][]{McKee82, Draine84}. In these models the population inversion is
primarily caused by absorption of $4.6~\mu$m photons that will excite
CO to the first vibrational level. From this level it decays back to
the ground vibrational state in higher rotational states. If the rate
of absorption of the $4.6~\mu$m photons is sufficiently high, an
inversion in the lowest rotational levels can occur. This requires the
$4.6~\mu$m lines to have relatively low optical depth so that the IR
emission can reach the region of the CSE where the population is not
collisionally quenched. A detailed analysis thus restricts possible
maser action to AGB stars that a have low mass-loss rate, a high CO abundance,
and a strong IR radiation field \citep{Morris80, Schoenberg88}. In
particular, it was concluded that the most likely candidate AGB stars
where CO maser action could be detected would be carbon stars with  low mass-loss rates
 \citep{Morris80}. Because the maser inversion is very
sensitive to the IR radiation, CO abundance, and the density in the
envelope, observations of CO masers could provide unique constraints
on localised regions of the CO envelope and on  the IR radiation
field coming from the stellar atmosphere.

\begin{figure*}[ht!]
\centering
\includegraphics[width=\textwidth]{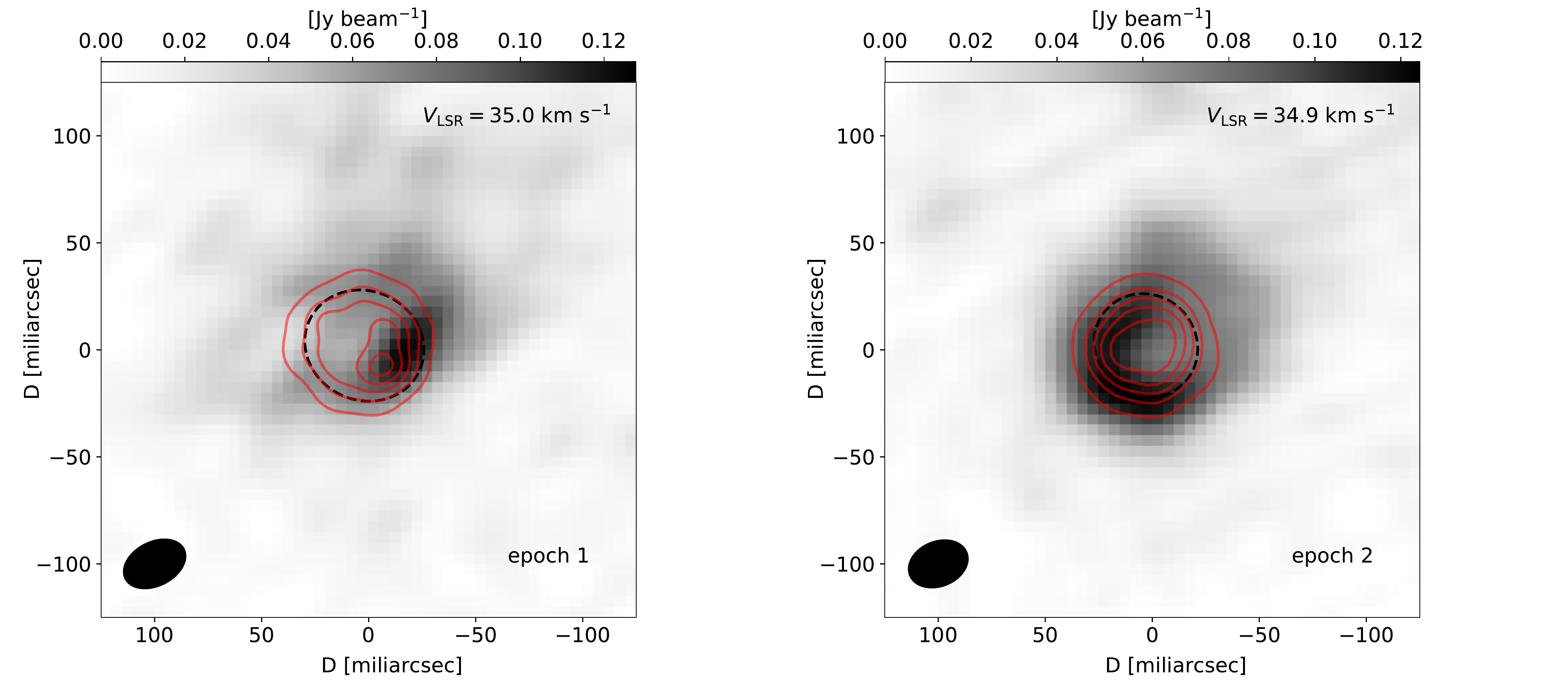}
\caption{Maps representing the channel with the brightest CO~$J=3-2$ emission around W~Hya for the two observational epochs (30 Nov--5 Dec 2015 and 25 Nov 2017). The strongest emission against the stellar continuum represents CO maser amplification. The grey scale is the continuum subtracted CO emission for a single channel $\sim1$~km~s$^{-1}$ in width. The synthesised beam is indicated in the bottom left corner and the channel velocity is indicated in the top right corner. The red solid contours indicate the uniformly imaged stellar continuum at $10, 30, 50, 70,$ and $90\%$ of the peak emission. The black dashed ellipses are the fitted stellar disc, as determined in \citet{Vlemmings17} and \citet{Vlemmings19}.}
     \label{WHya_COmap}
\end{figure*}

There have been no confirmed detections of CO maser
emission to date. Narrow $J=2-1$ CO emission around the carbon star V Hydrae
was previously thought to be due to maser inversion
\citep{Zuckerman86}. However, later observations revealed that the
narrow emission is identical in the $J=1-0$ line  and is more likely a
kinematic effect \citep{Zuckerman89}. There are also no clear
detections yet of CO maser emission from molecular clouds.

Here we present the first unambiguous detection of CO maser emission,
in the $J=3-2$ transition, around the oxygen-rich AGB star
W~Hya. W~Hya is a semi-regular variable star located at $\approx 100$~pc from the Sun
\citep{Vlemmings03,vanLeeuwen2007} with a mass-loss rate of
$\sim10^{-7}$~M$_\odot$~yr$^{-1}$ \citep[e.g.][]{Maercker08,Khouri14}
and a main-sequence mass, as constrained by oxygen isotopic ratios
between 1.0 and 1.5~M$_\odot$ \citep[][]{Khouri14,Danilovich2017}.
Continuum observations of the extended stellar atmosphere with ALMA
show that, at $345$~GHz, the star has a brightness temperature of
$\sim2700$~K and a size of $\sim5$~au \citep{Vlemmings19}. These
observations have also revealed strong variable surface hotspots with
brightness temperatures sometimes in excess of $50000$~K
\citep{Vlemmings17, Vlemmings19}.

The maser is detected in high angular resolution observations taken
at two epochs.
We present the observations in \S\ref{obs} and the results in
\S\ref{res}. We compare in \S\ref{disc} the results with simple
radiative transfer models that investigate  the dependence on the
continuum radiation field \citep[following][]{Morris80} as well as a
simplified implementation of the line overlap pumping described in
\citet{Piehler91}. Our conclusions are summarised in \S\ref{conc}.

\section{Observations and data reduction}
\label{obs}

The observations of W~Hya presented in this paper were previously  
presented in \citet{Vlemmings17} and \citet{Vlemmings19}. Here we
summarise the observation set-up and calibration steps. The first epoch
of observations of W~Hya in ALMA band 7, which included the CO~$J=3-2$
transition at $345.796$~GHz, were taken as part of ALMA project
2015.1.01446.S (PI: Takigawa) in three executions between 30 Nov  and
5 Dec  2015 for a total  on-source time of $\sim2.1$~hr. The second epoch was observed as part of project
2016.A.00029.S (PI: Vlemmings) on 25 Nov 2017 with an on-source observing time of $\sim35$~min.  All observations
were taken in spectral line mode, and the spectral window (spw) that
contained the CO~$J=3-2$ transition provided a native spectral
resolution of 0.98 MHz ($\sim0.9$~\kms) for both epochs. The initial
calibration (of the bandpass, amplitude, absolute flux-density scale,
and phase) was done using the Common Astronomy Software Applications
(CASA) package \citep{McMullin07}, using the 
manual calibration scripts provided by the observatory. Further observation details, including the
flux and gain calibrators used, are given in \citet{Vlemmings17} and
\citet{Vlemmings19}.

After the initial calibration we performed two rounds of phase
self-calibration on the strong stellar continuum. The self-calibration
improved the continuum signal-to-noise ratio (S/N) by a factor of three. Imaging of the CO
lines was done using Briggs weighting with a robust parameter of $-1.0$
and $0.5,$ respectively for the two  epochs, in order to achieve a similar
angular resolution. The resulting beam sizes were $31\times21$~mas
($-63^\circ$ from north) and $29\times21$~mas ($-68^\circ$). The maximum
recoverable scale is approximately $0.6\arcsec$. Despite the
additional observing time for epoch 1, the rms noise near the CO line
was $\sim3.5$~mJy~beam$^{-1}$ for both epochs due to the different
weighting adopted.

\section{Results}
\label{res}

\begin{figure*}[ht!]
\centering
\includegraphics[width=0.95\textwidth]{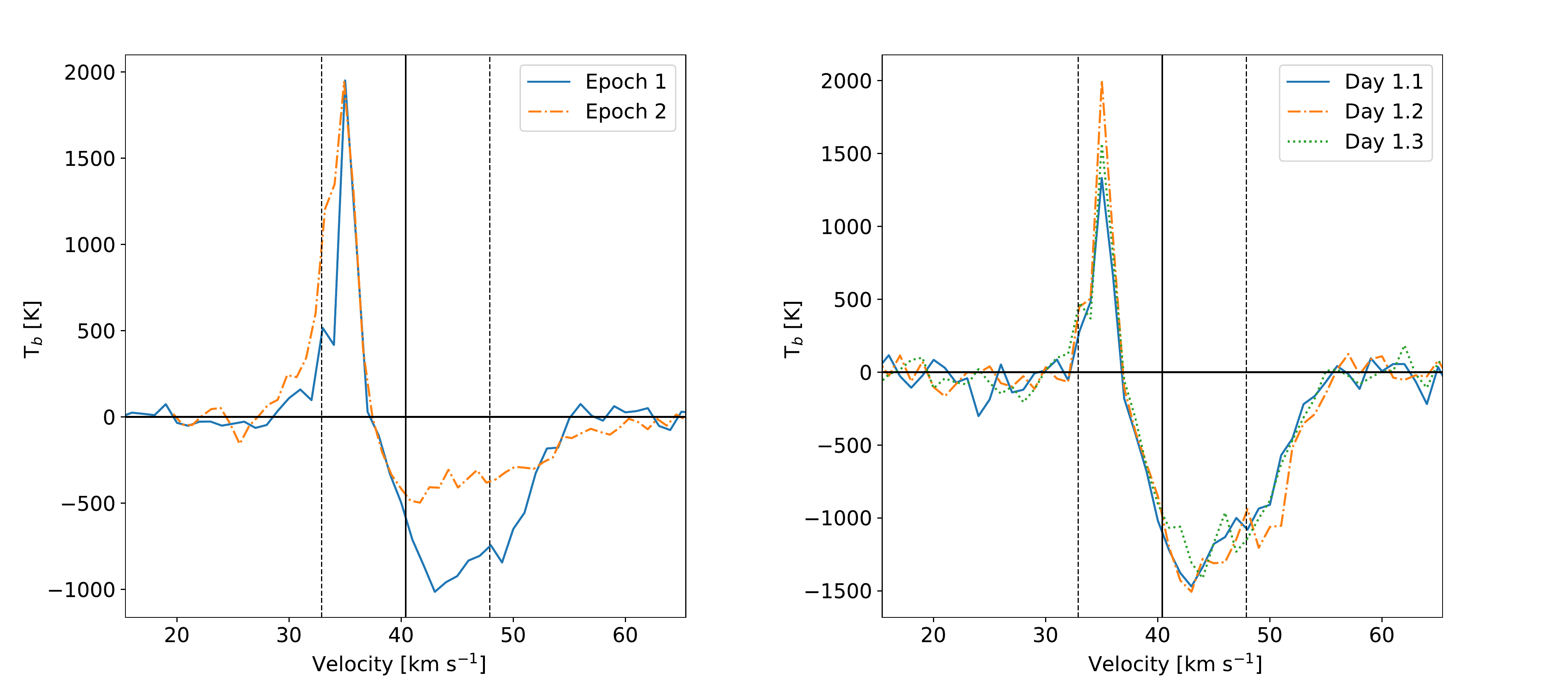}
\caption{Spectra of the CO~$J=3-2$ emission towars W~Hya. (left) The emission at the two
  observational epochs extracted from a single beam towards the
  peak of the maser emission. The vertical lines denote the
  stellar velocity $V_{\rm lsr}$ and the expansion velocity $V_{\inf}$
  of the envelope around W Hya \citep{Khouri14}. (right) Observations from the first epoch separated into the three observation days   (30 Nov, 3 Dec, and 5 Dec 2015) extracted at the location of the peak of the emission map of the combined days. The data were convolved to a common beam size of $30\times21$~mas.}
\label{COspec}
\end{figure*}

In both observational epochs, bright and relatively compact
CO~$J=3-2$
emission is detected. The emission maps of the brightest channels are
shown in Fig.~\ref{WHya_COmap}. After subtracting the $\sim2700$~K
stellar continuum, the peak of the emission has a brightness
temperature of $T_b\approx1950$~K in the two epochs. The emission
region is complex and, if it consists of multiple more compact
components, the derived $T_b$  presents a lower limit. As the
emission is amplifying the underlying stellar continuum, the
brightness temperatures imply a maser optical depth of
$\tau\approx-0.55$.

The morphology of the bright masing CO emission has markedly changed
between the two observing sessions. During the first epoch the
emission appears the be dominated by a relatively compact region
towards the western hemisphere of the star, close to the bright
hotspots observed on the stellar continuum surface
\citep{Vlemmings17}. In the second epoch the emission is more
extended and follows along the eastern limb of the stellar continuum.

The CO spectra taken in a single beam towards the peak of the maser
emission are shown in Fig. \ref{COspec}(left). In both epochs the maser
emission is confined to a few velocity channels ($\sim1$~km~s$^{-1}$ in width) between $33$ and $36.5$~km~s$^{-1}$ with a peak at $\sim35$~km~s$^{-1}$. The brightest spectral
component has a full width at half maximum (FWHM) of $1.9$ and $2.2$~km~s$^{-1}$
in the two epochs, respectively. The masers are superimposed on an
absorption profile originating from CO in the extended stellar
atmosphere. This absorption profile is of very different depth at
the position of the CO maser at each epoch. To investigate if there is
any short-term variability in the CO maser emission we also individually imaged
the three days in epoch 1 (30 Nov, 3 Dec, and 5 Dec). Although the synthesised beams are different, the
morphology of the maser emission does not change
significantly. The spectra of the three days, taken towards the peak
identified in the combined epoch 1 image and convolved to a common beam size, are shown in
Fig.~\ref{COspec}(right). The emission brightness temperature  appears to change from
$T_b\approx1330$~K on 30 Nov to $T_b\approx2000$~K on 3 Dec and to
$T_b\approx1570$~K on 5 Dec 2015. If the maser emission is more compact than the common beam of $30\times21$~mas, these brightness temperature values would correspond to lower limits. This indicates that there
could be variability of the order of $\pm20\%$ within only a five-day
timescale. For comparison, the continuum flux varies by less than
$2.5\%$ over the three days so the flux variation is not due to
an uncertainty in the flux calibration.

\section{Discussion}
\label{disc}

\subsection{Morphology and variability}
\label{morph}

\begin{figure*}[ht!]
\centering
\includegraphics[width=1.0\textwidth]{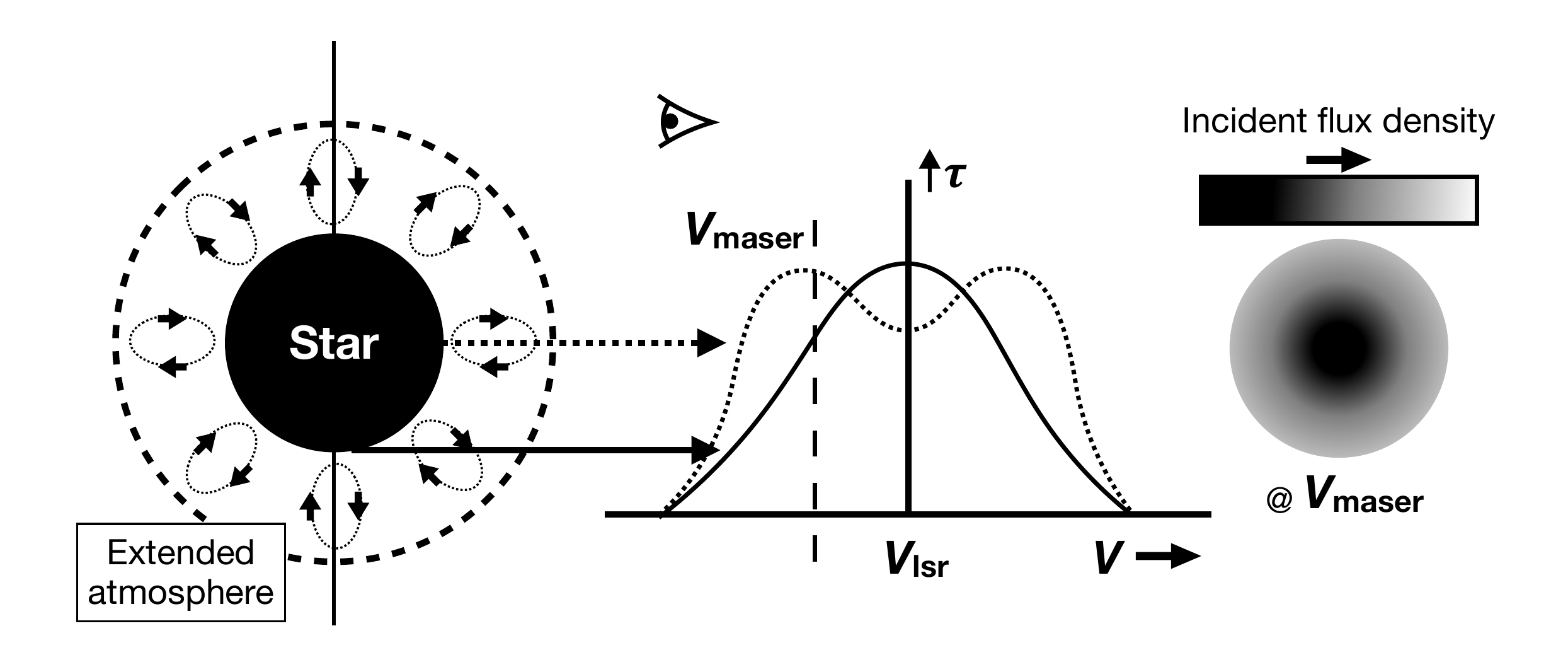}
\caption{Schematic illustrating the limb brightening of the emission
  incident on the circumstellar CO envelope amplified by the maser
  inversion of the foreground CO gas. The left panel illustrates the
  continuum emission from the star (solid circle) that is absorbed by
  the turbulent and inwards/outwards moving CO gas in the extended
  atmosphere. The dust formation zone is indicated by the dashed
  circle.  The dashed and solid lines represent the lines of sight
  towards the centre and limb of the star, respectively, for an
  observer on the right side of the page. The center panel illustrates
  the optical depth. Towards the stellar centre the optical depth
  ($\tau$) peaks towards high velocities representing the inward and
  outward motions of the gas in the extended atmosphere. In contrast,
  towards the limb of the star, the projection of the velocity vectors
  in the extended atmosphere result in the strongest absorption closer
  to the stellar systemic velocity. Therefore, the centre of the
  stellar disc is dimmer than the limbs at the maser velocities, as
  shown in the rightmost scheme. The brightest maser emission is thus
  expected towards the limb of the star where it amplifies the
  strongest incident emission.}
\label{schematic}
\end{figure*}

As seen in Fig.~\ref{WHya_COmap}, a common feature at the two epochs
is that the brightest CO maser emission is found projected towards the
limb of the stellar continuum. Although CO emission also
originates from the extended AGB atmosphere, maser inversion of the
ground state CO lines cannot occur there because the density of the
collision partners is too high \citep[e.g.][]{Vlemmings17,
  Khouri18}. Thus, the maser emission has to originate from CO gas
beyond the dust formation region where the wind is accelerated. There
the CO maser in front of the star will amplify the stellar continuum
emission. However, as illustrated in Fig.~\ref{schematic}, at the expansion velocity of the large-scale outflow, the CO molecules in the extended atmosphere absorb more efficiently towards the stellar centre than the limbs. Hence, the CO molecules in the outflow see a limb-brightened star, and amplification is naturally strongest towards the limbs. This can also explain the
morphological changes between the observing epochs as the maser
emission will depend on a combination of stellar surface emission
variations (such as the observed hotspots) and the depth of the
molecular absorption at the maser velocity. Finally, the maser
emission will also be affected by a possible density and velocity
structure in the foreground material. However, because the maser
inversion is small, this effect will be much less pronounced compared
to SiO, H$_2$O, and OH masers. Furthermore, transverse motions of dense
clumps across the full stellar disk within two years would, at the
distance of W~Hya, require a transverse velocity of
$\sim10$~km~s$^{-1}$, which is much higher than the turbulent
velocities estimated in CSEs ($\sim0.7$~km~s$^{-1}$ for W~Hya,
\citealt{Khouri14}). Thus, we suggest that the changes in the maser
emission primarily reflect changes in the underlying structure of the
star and extended atmosphere.

The flux variation of $\pm20\%$ observed over five days during the
first epoch is likely due to a combination of changes in the extended
atmosphere coupled with small changes in foreground masing
material. If the maser is amplifying the same background emission, a
$\pm20\%$ variability reflects a similar change in the background
emission or small variations in the maser optical depth between $\tau\sim-0.4$
and $-0.56$. The background continuum emission itself does not change
by more than a few percent, which means that it is unlikely that the observed variability is the result of rapid changes in the stellar continuum. As seen in Fig.~\ref{COspec}(right), the
atmospheric CO absorption itself shows that variations up to $\sim10\%$ could be present. The maser
optical depth is also very sensitive to the pumping emission at
4.6~$\mu$m and the effect of line overlap (see \S~\ref{models}). Thus, a small change in the background 4.6~$\mu$m
emission and CO absorption of the stellar continuum in the extended atmosphere, together with small-scale turbulence in
the masing gas that alters the coherent velocity path length, can
cause the observed variation in maser flux.

%----------------------------------------------------------------- 
\begin{figure*}[ht!]
\centering
\includegraphics[width=0.95\textwidth]{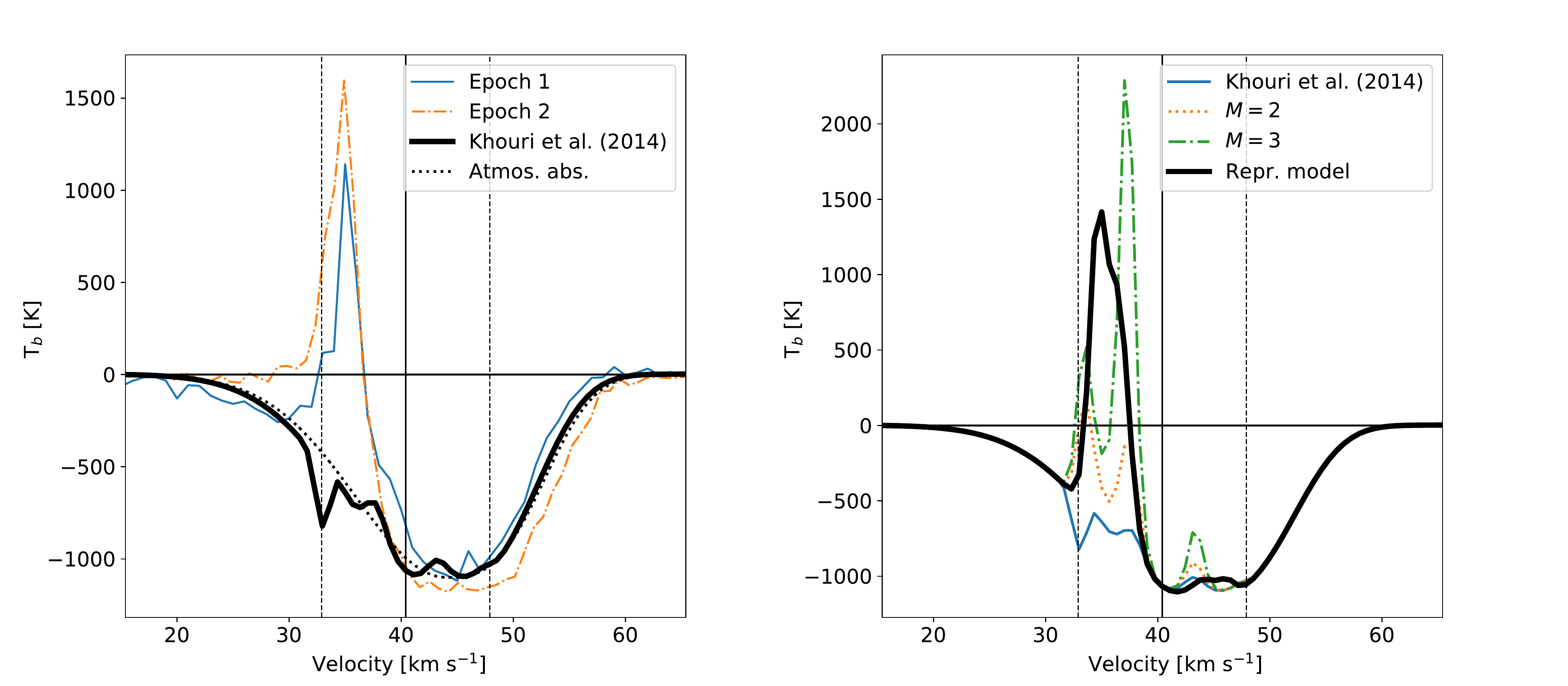}
\caption{Spectra of the CO~$J=3-2$ emission and the result of radiative transfer models. (left) The thin lines indicate the CO emission at the two
  observational epochs extracted using a $30$~mas aperture towards the
  stellar continuum. The thick line corresponds to the radiative
  transfer model from \citet{Khouri14}. The vertical lines indicate the stellar velocity and expansion velocity as in Fig.~\ref{COspec}. The absorption profile (black dotted line) 
  due to atmospheric CO was added to the model during
  ray-tracing. (right) Radiative transfer models of the CO~$J=3-2$
  emission around W~Hya including an enhanced IR radiation field with
  values of $M=2$ and $M=3$, compared to the baseline model with no
  IR enhancement from \citet{Khouri14} (equivalent to $M=1$) and a representative model that reproduces the
  observations (see text). The emission was taken from an aperture of
  $30$~mas. Because the radiative transfer model does not
  include the increased continuum opacity of the extended stellar
  atmosphere at submillimetre and radio wavelengths
  \citep[e.g.][]{Vlemmings19}, weak slightly red-shifted emission from
  behind the star is also seen in the model.}
\label{figspecs}
\end{figure*}

\subsection{Velocity}
\label{vel}

The velocity of the CO maser remained stable at $\sim35$~km~s$^{-1}$
over the two epochs with a FWHM velocity width of
$\sim2$~km~s$^{-1}$. This means that the maser coincides in velocity with
the brightest 1665 and 1667~MHz OH maser at $35.6$~km~s$^{-1}$,
observed between 1999 and 2002, which was shown to be amplifying the
background stellar radio continuum \citep{Vlemmings03}. At 183~GHz,
H$_2$O masers show a very distinct emission peak at
$35.1$~km~s$^{-1}$, with almost the same strength as the main peak
\citep{Humphreys17}. A similar feature, though  slightly
weaker compared to the peak emission, is seen in the 1.296~THz H$_2$O
maser \citep{Neufeld17}, while observations of the 658~GHz H$_2$O
maser show a weak but distinct peak at the same velocity
\citep{Baudry18}. The 321 and 325~GHz H$_2$O maser spectra from
\citet{Yates95}  show multiple peaks, but they do not extend to the
velocity where we find the CO maser. Similarly, the 22~GHz H$_2$O
masers spectra show no emission at around $35$~km~s$^{-1}$ \citep{Imai19}
and imaging with the MERLIN interferometer seems to show that the 22~GHz
masers are predominantly found in a tangentially amplified shell
between $\sim5$ and $25$~au from the star \citep{Richards12}.

Even if all features seen in the various masers around
$35$~km~s$^{-1}$ are related, it is unlikely that this is due to a
physical association in a single maser cloud. The 658~GHz H$_2$O
masers require a high temperature ($>1000$~K) and H$_2$O density
($>10^5$~cm$^{-3}$), while the 183~GHz H$_2$ masers are found over a
larger range of temperatures (between $\sim400$~K and $2500$~K), but
with an H$_2$O density $<10^5$~cm$^{-3}$. The conditions for the
1.296~THz masers fall roughly  between the other two maser
transitions \citep{Gray16}. Considering an H$_2$O abundance of
$1.5\times10^{-3}$ for W~Hya \citep{Maercker08}, this means that the
H$_2$O masers are excited at molecular hydrogen densities of $n_{\rm
  H_2}\sim 10^{6}$ up to $10^9$~cm$^{-3}$. For the typically assumed
mass-loss rate of W~Hya this corresponds to a region within
$\sim35$~au, and hence well within the region where we would expect the
amplified stellar image in the OH masers. According to an OH maser
shell model of \citet{Szymczak98}, the OH masers occur in a
shell with a radius of $\sim1000$~au and a thickness of $\sim300$~au,
although the fit of the observations to a thin shell is
poor. Furthermore, the time baseline of the various observations spans
more than 20~years. It is thus more likely that the velocity of the CO
maser and the other features corresponds to the largest coherent
velocity path length towards the star, which results in an amplified
stellar image in a large number of different maser species and
transitions.

%----------------------------------------------------------------- 
\begin{figure*}[ht!]
\centering
\includegraphics[width=0.95\textwidth]{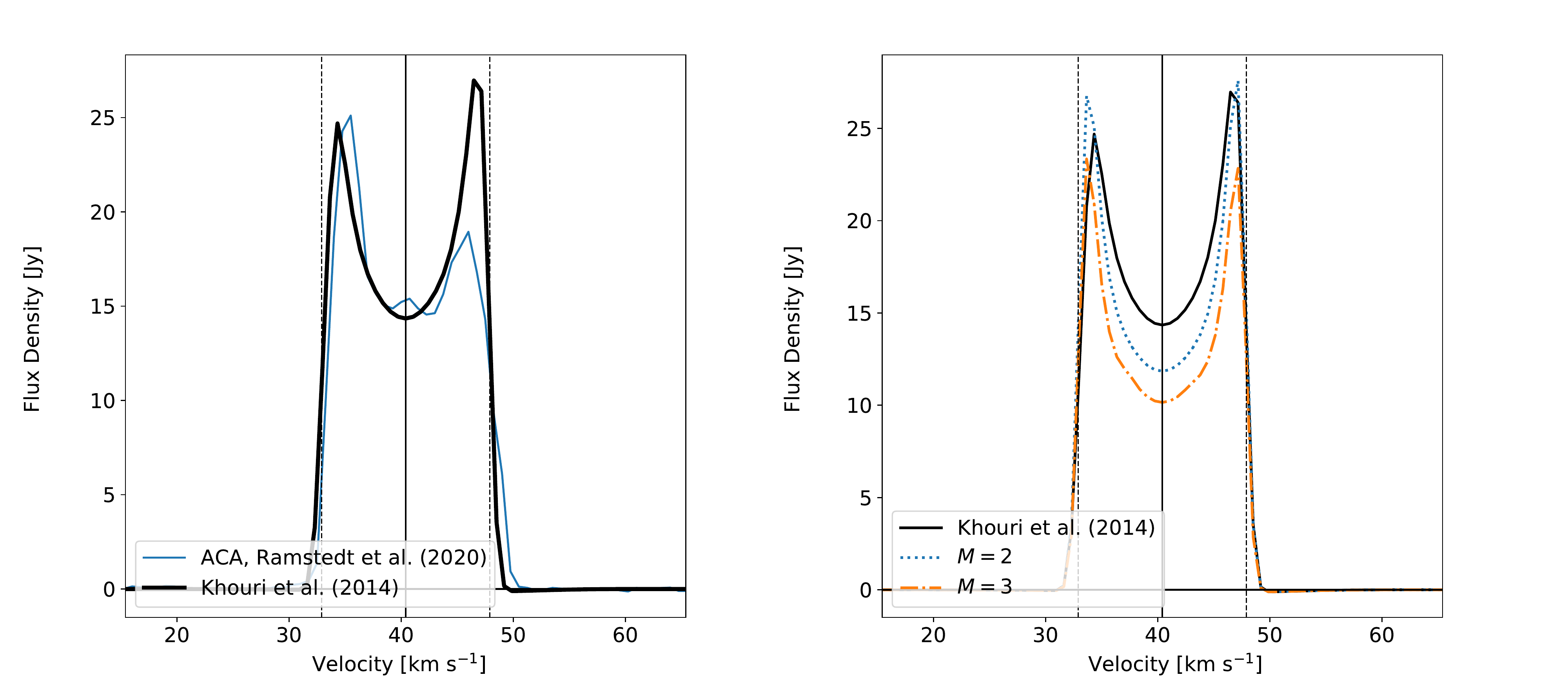}
\caption{Spectrum and radiative transfer models of CO~$J=3-2$ emission at ACA resolution ($4.5\times2.9\arcsec$) (left) Spectrum of the emission around W~Hya as
  observed with the ACA \citep[project 2016.2.00025.S,][]{Ramstedt20} extracted in a single beam
  towards the star.  The black line
  corresponds to the radiative transfer model from \citet{Khouri14}
  shown for a small pencil beam in Fig.~\ref{figspecs}(left). The
  vertical lines correspond to the same lines in
  Fig.~\ref{COspec}. (right) Radiative transfer models of the
  CO emission including an enhanced
  IR radiation field with values of $M=2$ and $M=3$, compared to the
  baseline model with no IR enhancement $M=1$.}
\label{ACAspecs}
\end{figure*}

\begin{figure}[ht!]
\centering
\includegraphics[width=0.5\textwidth]{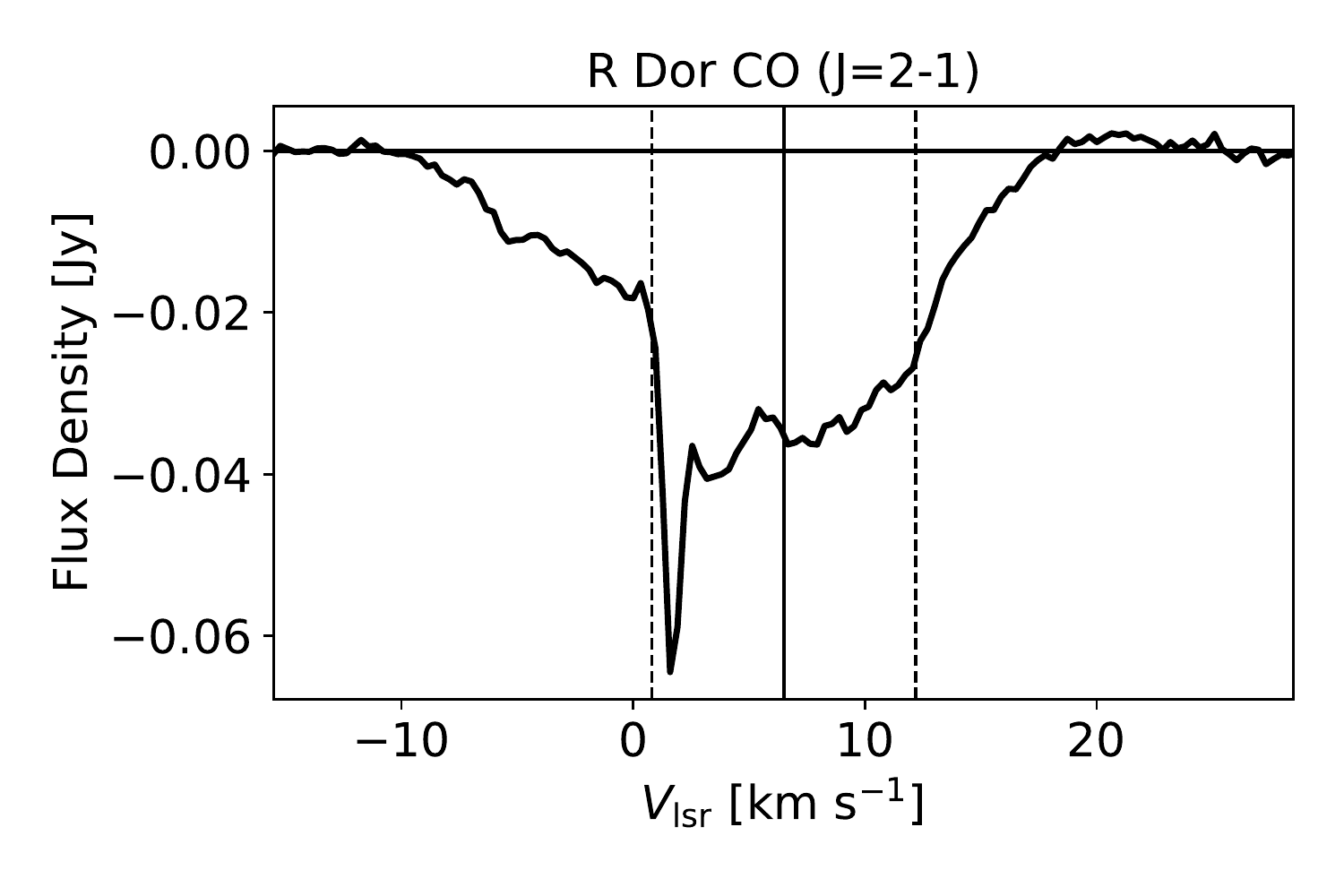}
\caption{Spectrum of the CO~$J=2-1$ absorption and emission around R~Dor extracted in a $45$~mas aperture
  towards the star from ALMA project 2017.1.00824.S (PI Decin). The vertical lines
  correspond to the stellar velocity (solid line) and expansion velocity $V_{\inf}$ (dashed lines)
  derived from molecular line observations \citep{DeBeck2018}. Despite a similar mass-loss rate to W~Hya ($\sim1\times10^{-7}~\dot{\rm M}$~yr$^{-1}$) no maser is detected in the CO~$J=2-1$ transition.}
\label{RDor}
\end{figure}

\begin{figure}[ht!]
\centering
\includegraphics[width=0.5\textwidth]{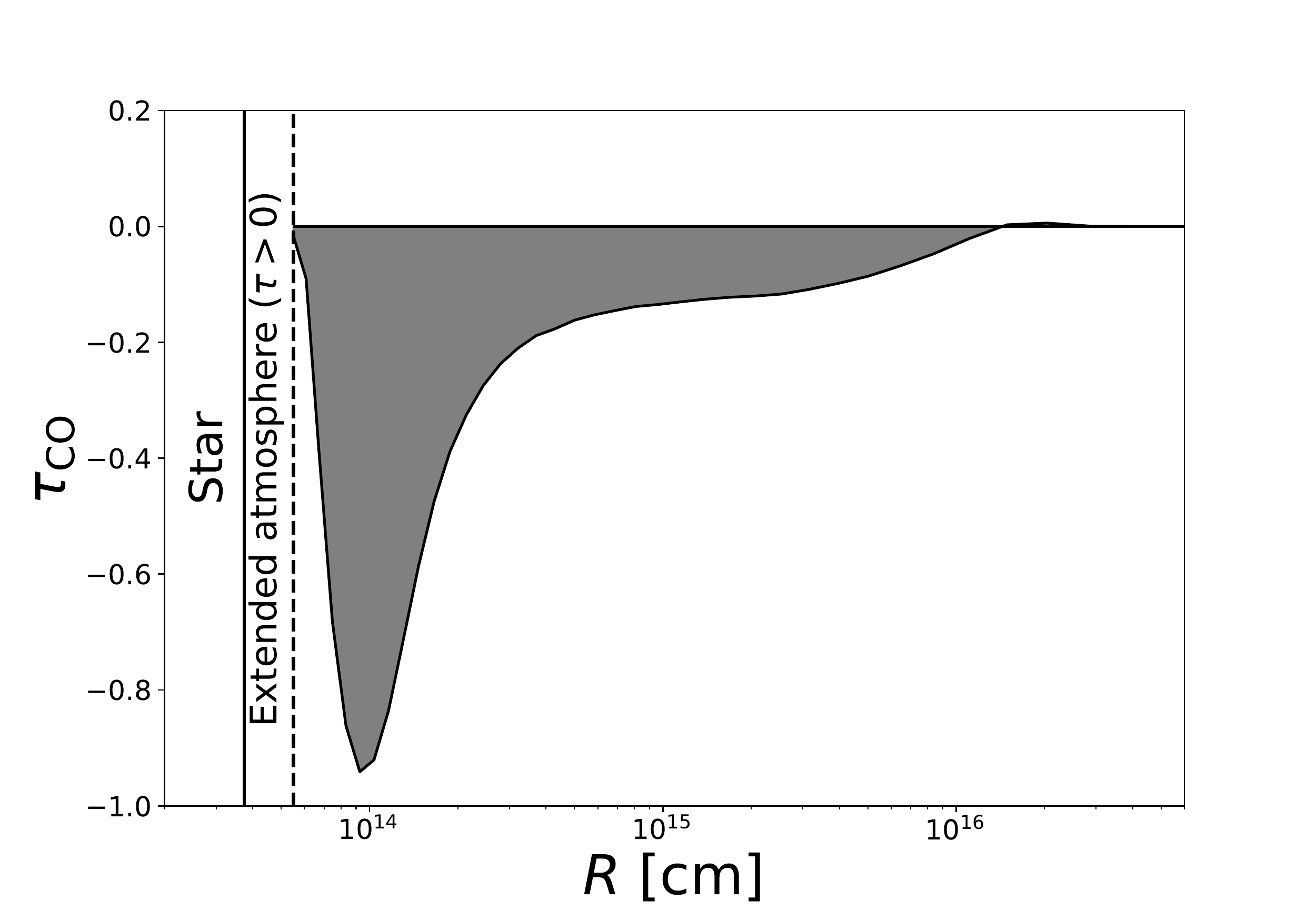}
\caption{Optical depth in the CO~$J=3-2$ line along the radial line of sight to the star for the representative model shown in Fig.~\ref{figspecs}. For this model the negative optical depth peaks at $\sim7$~au. Weak population inversion exists throughout almost the entire circumstellar envelope. Also shown are the stellar radius and the size of the extended stellar atmosphere.}
\label{tau}
\end{figure}

\begin{figure*}[ht!]
\centering
\includegraphics[width=0.95\textwidth]{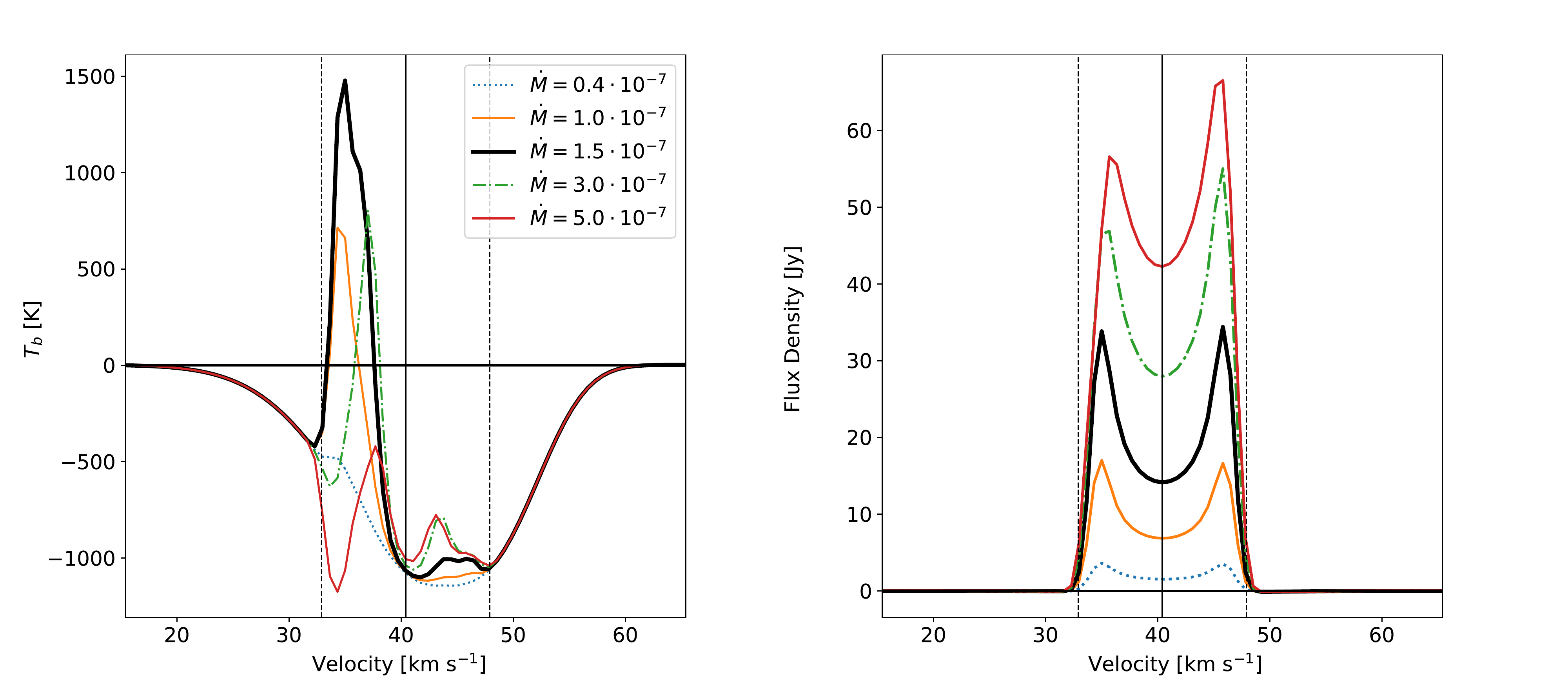}
\caption{As in Fig.~\ref{figspecs} and Fig.~\ref{ACAspecs}. Here, the left panel presents the most representative maser model described in \S~\ref{models} for a range of stellar mass-loss rates. The right panel shows the model spectra as they would be observed with the ACA.}
\label{MLspecs}
\end{figure*}

\subsection{Radiative transfer models}
\subsubsection{Continuum pumping}
\label{models}

\begin{table}
\caption{Adopted parameters of the baseline CO model$^a$}             
\label{Table:COmod}      
\centering          
\begin{tabular}{r r }     % 2 columns 
\hline\hline       
\multicolumn{2}{l}{Stellar parameters:}\\
T$_*$ & 2500~K \\
L$_*$ & 5400~L$_\odot$ \\
D & 115~pc\\
\hline
\multicolumn{2}{l}{Envelope parameters:}\\
$\dot{\rm M}$ & $1.5\times 10^{-7}$~M$_\odot$~yr$^{-1}$ \\
$R_{\rm in}$ & 3.68~au \\
\hline
\multicolumn{2}{l}{Abundance and e-folding radius:}\\
$X_{\rm CO}$ & $4\times10^{-4}$ \\
$R_{\rm 0.5}$ & 630~au \\
\hline
\multicolumn{2}{l}{Gas velocity law:}\\
\multicolumn{2}{l}{$V(r)=V_{\rm 0}+(V_{\infty}-V_{\rm 0})(1-R_{\rm in}/r)^{\beta}$}\\
$\beta$ & 5.0\\
$V_{\rm 0}$ & 3.0\\
$V_{\infty}$ & 7.5\\
\multicolumn{2}{l}{Turbulent velocity:}\\
$V_t$ & 0.8 \\
\hline
\multicolumn{2}{l}{Dust:}\\
$\tau_{10 \mu{\rm m}}$ & 0.1 \\
Type & Silicate\\
\hline\hline       
\end{tabular}
\tablefoot{
\tablefoottext{a}{From \citet{Khouri14}}
}
\end{table}

In order to investigate the occurrence of the CO maser in the $J=3-2$
transition, we perform one-dimensional radiative transfer models using
the Monte Carlo radiative transfer code {\it mcp}
\citep{Schoeier01}. We include the CO rotational levels $J=0$ to
$J=81$ and the vibrational levels $v=0$ and $v=1$. The transition frequencies, energy levels, and Einstein coefficients were taken from the CDMS and JPL databases \citep{Muller01, Picket98} and the collisional rate coefficients come from \citep{Yang10}.

It has long been
recognised that mild population inversion of CO can occur for a narrow
mass-loss range \citep[e.g.][]{Morris80}. The maser range is set by the
mass-loss rate (or gas density) and CO abundance as well as the
availability of 4.6~$\mu$m pump photons. At high gas density, the CO
is mostly collisionally excited, quenching the maser. A high CO column
density also produces a high opacity at the  4.6~$\mu$m 
CO~$v=0-1$ transitions reducing the volume over which the maser
can operate. Conversely, as the maser requires a sufficiently large
negative optical depth, a  CO column density that is too low also reduces the
maser strength. To investigate the occurrence of the CO maser around
W~Hya, we model the observations of a single aperture (with a size of
$2R_*$) towards the star using the parameters for the CO envelope
given in Table \ref{Table:COmod} taken from \citet{Khouri14}; we do not adjust the distance and adopt the value used in the
model. Since the model is one-dimensional we only compare the
emission along the line of sight to the star.  In order to include the
effect of the broad CO absorption that originates from the extended
atmosphere, we include this absorption profile on top of the stellar
black body when ray-tracing the solutions of the radiative transfer
model. We include a comparison of the model with recent ACA
observations (with a beam size of $4.5\times3.6\arcsec$). The results of this model are shown in
Fig.~\ref{figspecs}(left). The ACA spectrum and corresponding model
are shown in Fig.~\ref{ACAspecs}(left). No maser emission is generated
with these model parameters.  Additionally, at the terminal velocity,
an absorption component is seen that is the result of observing towards the star through the cool blue-shifted CO envelope in the
foreground. Such a narrow absorption component is absent in our observed
spectrum, although we do see this effect in high angular resolution
observations of the CO~$J=2-1$ emission around the nearby AGB star
R~Dor (shown in Fig.~\ref{RDor}).

Subsequently, we artificially adjust the strength of the 4.6~$\mu$m
radiation field, following the same method as in \citet{Morris80}, to
investigate when maser emission is generated. Specifically, we
multiply the radiation field for all wavelengths $<5~\mu$m by a factor
$M$. The result of these models are shown in
Figs.~\ref{figspecs}(right) and \ref{ACAspecs}(right).  Several things
become apparent from these models. Firstly, a maser of sufficient
strength can be generated if the radiation field at 4.6~$\mu$m is
increased by a factor $M\sim3$. However, because the inversion mostly
occurs in the inner envelope, the maser velocity is too close to the
stellar velocity in the adopted model. Finally, the maser width is too
narrow. In order to overcome these issues, we require a larger
velocity at the inner envelope ($V_0\approx4.2$~km~s$^{-1}$), a
slightly steeper acceleration ($\beta\approx 2.5$), and larger
turbulent velocity component ($V_t\approx 1.0$~km~s$^{-1}$). Varying
these three parameters, we can reproduce the maser as seen in the
representative model in Fig.~\ref{figspecs}(right). While making these
changes, we did not consider further constraints from the lower
angular resolution observations. In order to investigate where, in this model, most of the maser emission originates with respect to the star, the extended atmosphere, and the CSE, we present the radial CO optical depth $\tau$ as a function of radius in Fig.~\ref{tau}. Population inversion in the $J=3-2$ line persists throughout most of the CSE and peaks around $7$~au. This can be compared to the stellar radius (at 345~GHz) of $2.5$~au and the approximate size of the extended stellar atmosphere and start of the dust formation zone at $\sim 3.7$~au. In the extended atmosphere, no maser emission can occur because of the high densities. As discussed in \S~\ref{morph} and illustrated in Fig.~\ref{schematic}, the stellar continuum emission is absorbed by CO gas in the extended atmosphere before it is subsequently amplified by the foreground CO maser. The exact shape of the optical depth profile is strongly dependent on the velocity profile and enhanced IR radiation field adopted in the model.

In Fig.~\ref{MLspecs} we investigate the effect of changing the mass-loss rate on our maser model. Under the assumed conditions we find that maser emission is generated, from different parts of the envelope, for mass-loss rates between $0.5-5\times10^{-7}$~M$_\odot$~yr$^{-1}$. Specifically for W~Hya, most of this range can be ruled out by the lower angular resolution observations, limiting the rate to between $1.5-3.0\times10^{-7}$~M$_\odot$~yr$^{-1}$.

In our models we can reproduce the required continuum excess by including a dust shell with an optical depth at 4.6~$\mu$m of $\sim0.2$, which is located closer to the star than the maser-emission region. Such a dust shell would have optical depths greater by at least a factor of five at $\sim 10~\mu$m and $\sim 1~\mu$m, affecting the spectral energy distribution in a way that is not supported by the observed infrared spectrum \citep[e.g.][]{Justtanont14}. Although the model results show that the maser can be reproduced by a general increase in IR radiation field and fine-tuning the model parameters, we  conclude that this is not a viable solution. We thus
investigate pumping that can occur due to line overlap around
4.6~$\mu$m.

\begin{figure*}[ht!]
\centering
\includegraphics[width=0.95\textwidth]{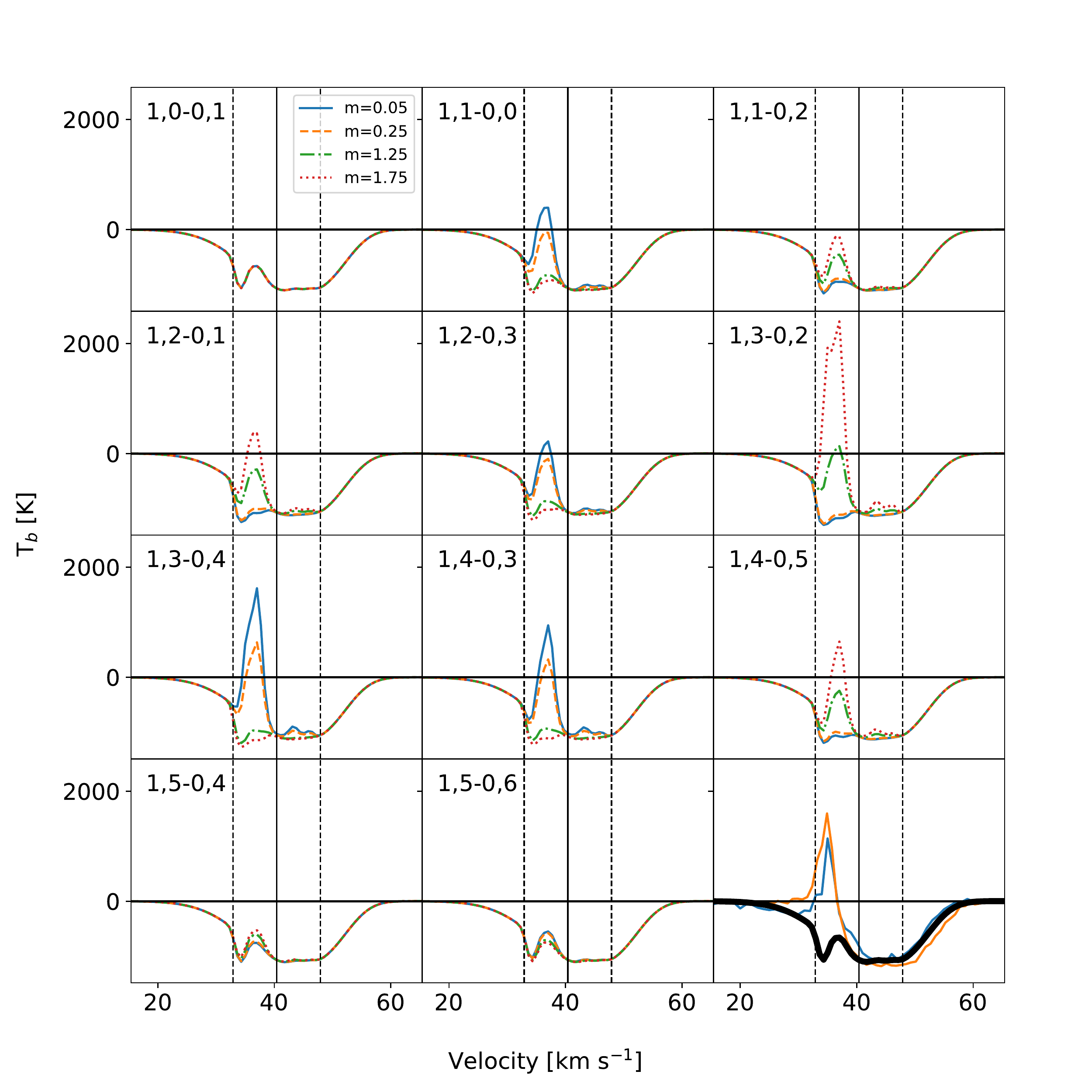}
\caption{Radiative transfer models of the CO~$J=3-2$ emission around W~Hya including increased or decreased IR pumping for selected $v=0$ to $v=1$ vibrational transitions around a wavelength of $4.6~\mu$m. The panels are labelled with the $v,J-v',J'$ quantum numbers of the transition at which the incident radiation field from the star is decreased or enhanced. The bottom right panel indicates the baseline model and the observations. Based on observations by \citet{Justtanont14}, the models assume that the radiation field at $4.6\mu$m for all transitions is reduced to 85\% of the stellar black-body radiation field due to atmospheric absorption. The incident black-body IR emission for the individual lines presented in each panel is multiplied by a factor $m$, where $m<0.85$ indicated enhanced absorption, while $m>0.85$ indicates a reduction in atmospheric absorption or even extra emission. The vertical lines are as in Fig.~\ref{figspecs}.}
\label{IRdiffspecs}
\end{figure*}

\subsubsection{Line overlap}
\label{overlap}

Line overlap of spectral lines around the pump wavelength at
4.6~$\mu$m has also been suggested as   the cause of CO maser
emission \citep{Piehler91}. Specifically, a
combination of absorption and emission lines originating from the
stellar atmosphere, can change the relative pumping of the
$v=0-1$ IR lines. Here we focus on the effect on the IR
lines of the $\Delta J\pm1$ transitions between the ground state and the first
vibrationally excited state for $J=0$ up to $J=5$. By relative changes
of the radiation field seen by these lines, maser inversion can
readily be obtained. In Fig.~\ref{IRdiffspecs} we show the results of changing the stellar radiation field experienced by these transitions. We adopted the CO envelope parameters from Table~\ref{Table:COmod}
with the adjustment to the velocity field as discussed previously. Based on observations around 4.6~$\mu$m \citep{Justtanont14} we include an overall decrease in the stellar radiation field around this wavelength, adopting a transmission of 85\% of the stellar black body.
As expected, the appearance of the CO~$v=0, J=3-2$ maser is closely related to the relative pumping into and out of the $J=3$ level. The maser emission in this transition requires a slight overpopulation of the $v=0, J=3$ level. This means that an increase in the intensity of the radiation field affecting the transition between the $v=1, J=3$ level and the $v=0, J=2$ levels will decrease the $v=0, J=2$ population in favour of the $v=1, J=3$ level. As a consequence, the ratio of $v=0, J=3$ to $v=0, J=2$ inverts. Similarly, a decrease in the intensity of the radiation field affecting the transition between $v=1, J=3$ and $v=0, J=4$ will reduce the pumping from the $v=0, J=4$ to the $v=1, J=3$ level. The subsequent decay of the larger population in the $v=0, J=4$ level to the $v=0, J=3$ level will result in maser inversion in the CO~$v=0, J=3-2$ transition. Our models and the more extensive study involving many spectral components in \citet{Piehler91} highlight that the maser emission is highly sensitive to the exact profile, including absorption and emission lines, of the stellar atmosphere. The current observations of a single transition can be reproduced by many different combinations of relative pumping in the IR. Still, considering the rich structure known to exist in the spectra originating from the stellar atmosphere, it is highly likely that the line overlap is the main reason for the observed CO maser. In particular, also because pumping due to continuum emission requires an increase in the IR radiation field that is not observed (see \S~\ref{models}). The range of mass-loss over which a maser can be reproduced using the pumping of a single IR line is similar to that shown in Fig. \ref{MLspecs} for the excess continuum IR emission. However, a detailed study of this would involve too many degenerate parameters to be constrained with a single observed maser transition.

A study involving multiple masing transitions can provide stronger constraints on the contribution of the different IR lines. In Appendix~\ref{others} we investigate the effect of selective pumping of the different IR lines on the CO~$J=2-1$ (Fig. \ref{IRdiffspecs21}) and $J=4-3$ (Fig.~\ref{IRdiffspecs43}) transitions. We also include the CO~$J=3-2$ results without the atmospheric absorption profile in Fig.~\ref{IRdiffspecs32}. These model results show that in particular the combination of these three lines can be used to investigate which $v=0-1$ IR transitions are responsible for the CO maser pumping since all three are affected differently by pumping the different IR lines. For example, the extra absorption between the $v=0, J=4$ and $v=1, J=3$ levels (which can explain the CO~$J=3-2$ maser of W~Hya) would not produce any maser in the $J=2-1$ transition. This could explain why no maser is observed in the CO~$J=2-1$ transition of R~Dor (Fig.~\ref{RDor}) even though R~Dor has a mass-loss rate similar to that of W~Hya.

\begin{figure*}[ht!]
\centering
\includegraphics[width=0.95\textwidth]{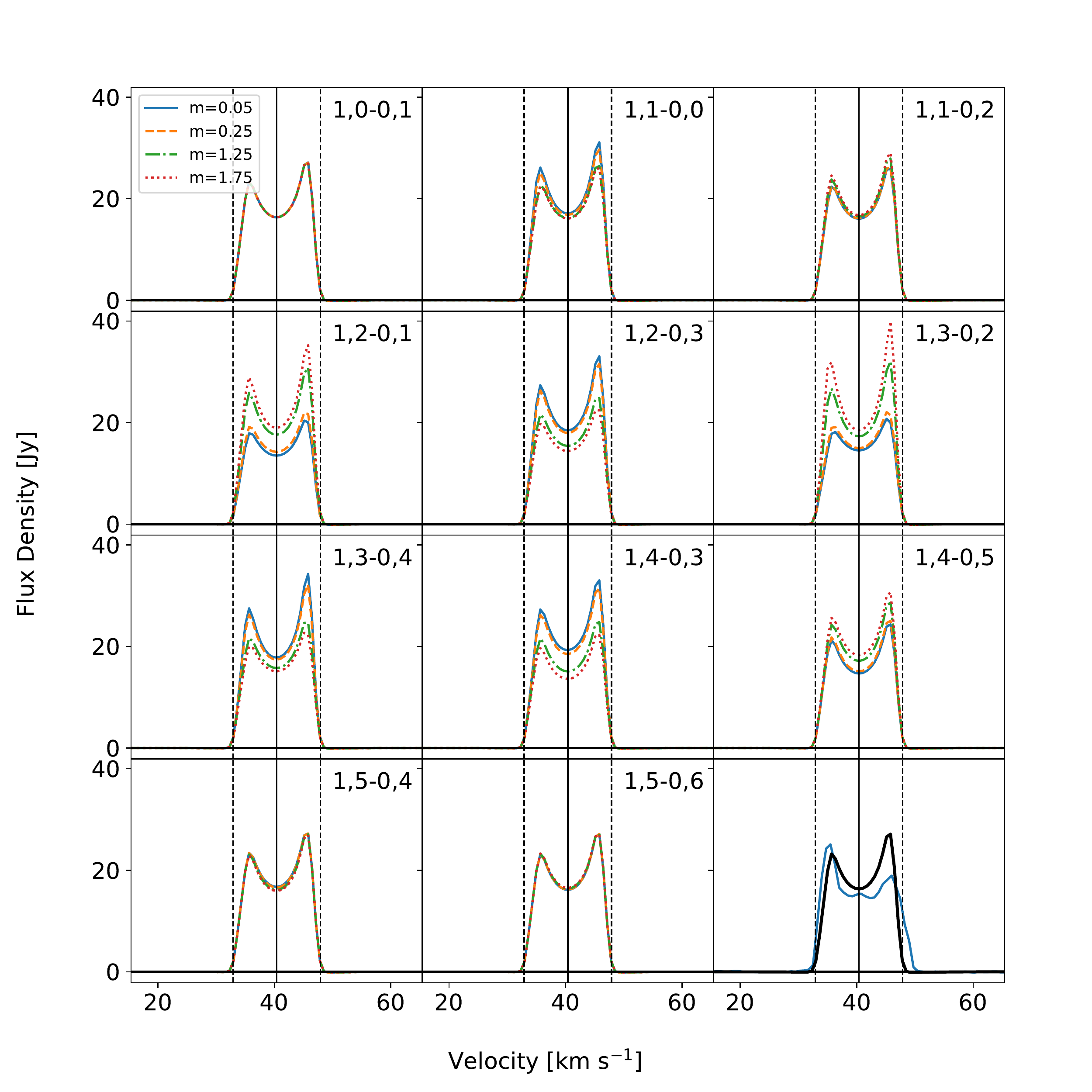}
\caption{As in Fig.~\ref{IRdiffspecs}, but  for the emission as observed with the ACA.}
\label{IRdiffspecsACA}
\end{figure*}

Similar to the continuum pumping, a change in the relative emission incident on the individual 4.6~$\mu$m CO transitions also changes the CO emission profile on the larger scales   shown in Fig.~\ref{IRdiffspecsACA}. Because of the large number of degeneracies  including the individual line pumping, we did not attempt to adjust the standard CSE model to simultaneously reproduce the larger scale observations. However, it is clear that the models that can reproduce the maser emission also produce changes in the larger scale emission that do not match the observations. The selective pumping our model requires has a different relative effect on each of the low-excitation rotational CO lines. Hence, model parameters such as mass-loss rate, velocity structure, and turbulent velocity will be much better determined when considering constraints from CO maser emission, when present.

\section{Conclusions}
\label{conc}

Using ALMA long baseline observations, we have confirmed that mild
maser inversion occurs in the CSE of the AGB star W~Hya. The maser
emission is shown to occur at a stable $V_{\rm LSR}$ velocity in two
observational epochs separated by two years. While the flux density of
the maser emission is also similar, the morphology has altered
significantly in this time. Since the maser is seen to amplify the
underlying stellar continuum, the changes in morphology are likely due
to a combination of small-scale structure in the masing material and
changes in the underlying stellar continuum. The maser emission is
found at $V_{\rm LSR}\approx 35$~km~s$^{-1}$, which corresponds,
within the spectral resolution, to the velocity of the amplified
stellar emission feature seen in the OH 1665 and 1667~MHz masers
\citep{Vlemmings03}. It also corresponds to the velocity of a bright
emission peak in the 183~GHz H$_2$O masers \citep{Humphreys17} and a
weaker feature in the 658~GHz and 1.296~THz H$_2$O masers \citep{Baudry18, Neufeld17}. This
implies a large column of material at coherent velocity in front of
the star since the excitation conditions of the masers and the CO
stretch over a wide range of densities and temperatures.  

We can reproduce the CO maser emission assuming a mass-loss rate
similar to that typically derived for W~Hya
($\sim10^{-7}$~M$_\odot$~yr$^{-1}$, e.g. \citealt{Khouri14}). We
find the CO~$J=3-2$ line can be inverted if the 4.6~$\mu$m continuum
radiation field for the circumstellar dust is significantly
enhanced. However, this requires a radiation field at 4.6~$\mu$m that is
inconsistent with observations \citep[e.g.][]{Justtanont14}. Instead,
the likely explanation for the maser emission involves a complex
structure of atmospheric emission and absorption lines that produce
differential pumping of the IR CO transitions around 4.6~$\mu$m,
between the ground level  and the  first vibrational level.  A more exact
determination of the relative pumping of these lines is not possible
with the current observations of only one CO maser transition and
would require near-simultaneous observations at high angular
resolution of multiple rotational transitions. Our models show that
the relative IR pumping that gives rise to the maser also introduces
an overall uncertainty in the molecular line radiative transfer
modelling for stars such as W~Hya. A proper inclusion of the IR lines
in the AGB atmosphere is thus needed to further reduce uncertainties
in mass-loss and radiative transfer modelling.

\begin{acknowledgements}
  WV and TK acknowledge support from the Swedish Research
  Council under grants No. 2014-05713 and 2019-03777. This paper makes use of the following ALMA data: ADS/JAO.ALMA\#2015.1.01446.S,
  ADS/JAO.ALMA\#2016.A.00029.S, ADS/JAO.ALMA\#2016.2.00025.S, and ADS/JAO.ALMA\#2017.1.00824.S. ALMA is a partnership of ESO
  (representing its member states), NSF (USA) and NINS (Japan),
  together with NRC (Canada), NSC and ASIAA (Taiwan), and KASI
  (Republic of Korea), in cooperation with the Republic of Chile. The
  Joint ALMA Observatory is operated by ESO, AUI/NRAO and NAOJ. We
  also acknowledge support from the Nordic ALMA Regional Centre (ARC)
  node based at Onsala Space Observatory. The Nordic ARC node is
  funded through Swedish Research Council grant No 2017-00648.
\end{acknowledgements}

% WARNING
%-------------------------------------------------------------------
% Please note that we have included the references to the file aa.dem in
% order to compile it, but we ask you to:
%
% - use BibTeX with the regular commands:
%   \bibliographystyle{aa} % style aa.bst
%   \bibliography{Yourfile} % your references Yourfile.bib
%
% - join the .bib files when you upload your source files
%-------------------------------------------------------------------

\bibliographystyle{aa}
%\bibliography{COmasers}

\begin{appendix}
\section{Other CO maser transitions}
\label{others}
\begin{figure*}[ht!]
\centering
\includegraphics[width=0.95\textwidth]{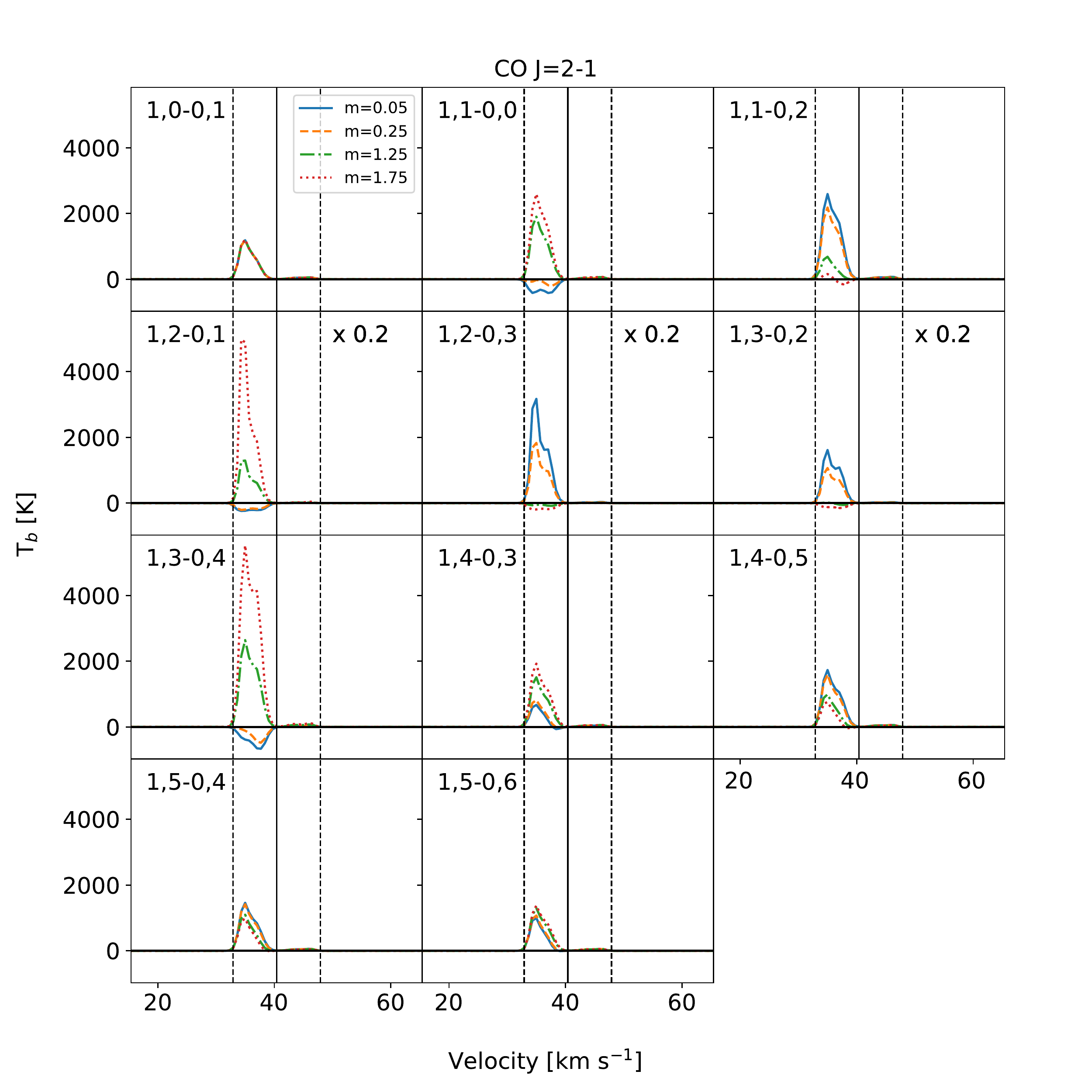}
\caption{As in Fig.~\ref{IRdiffspecs}, but without the atmospheric absorption component for the CO~$J=2-1$ emission observed towards the star with a 30~mas aperture. The spectra are scaled by a factor of five in the panels of the second row.}
\label{IRdiffspecs21}
\end{figure*}

\begin{figure*}[ht!]
\centering
\includegraphics[width=0.95\textwidth]{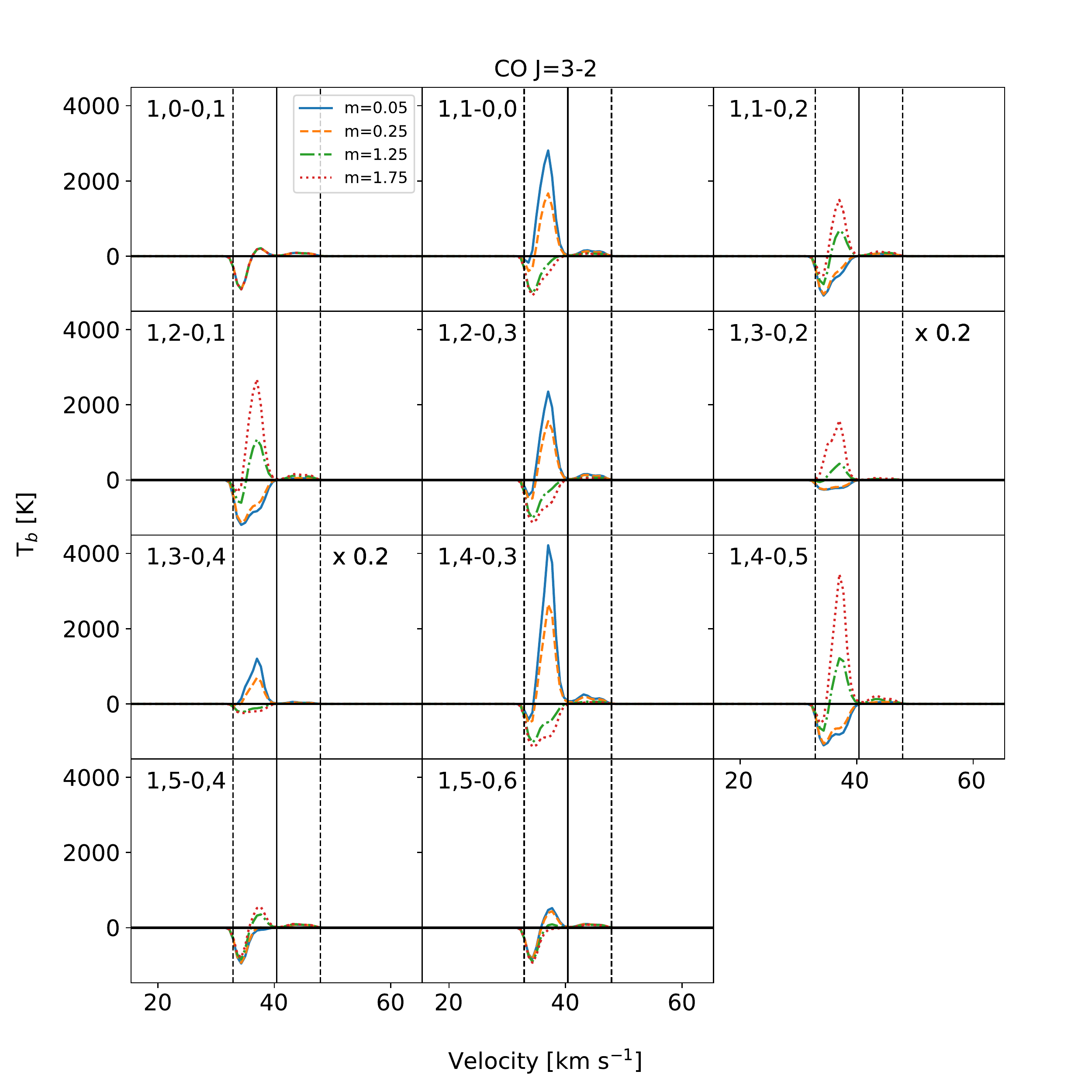}
\caption{As in Fig.~\ref{IRdiffspecs}, but without the atmospheric absorption component for the CO~$J=3-2$ emission observed towards the star with a 30~mas aperture. The spectra are scaled by a factor of five in two panels.}
\label{IRdiffspecs32}
\end{figure*}

\begin{figure*}[ht!]
\centering
\includegraphics[width=0.95\textwidth]{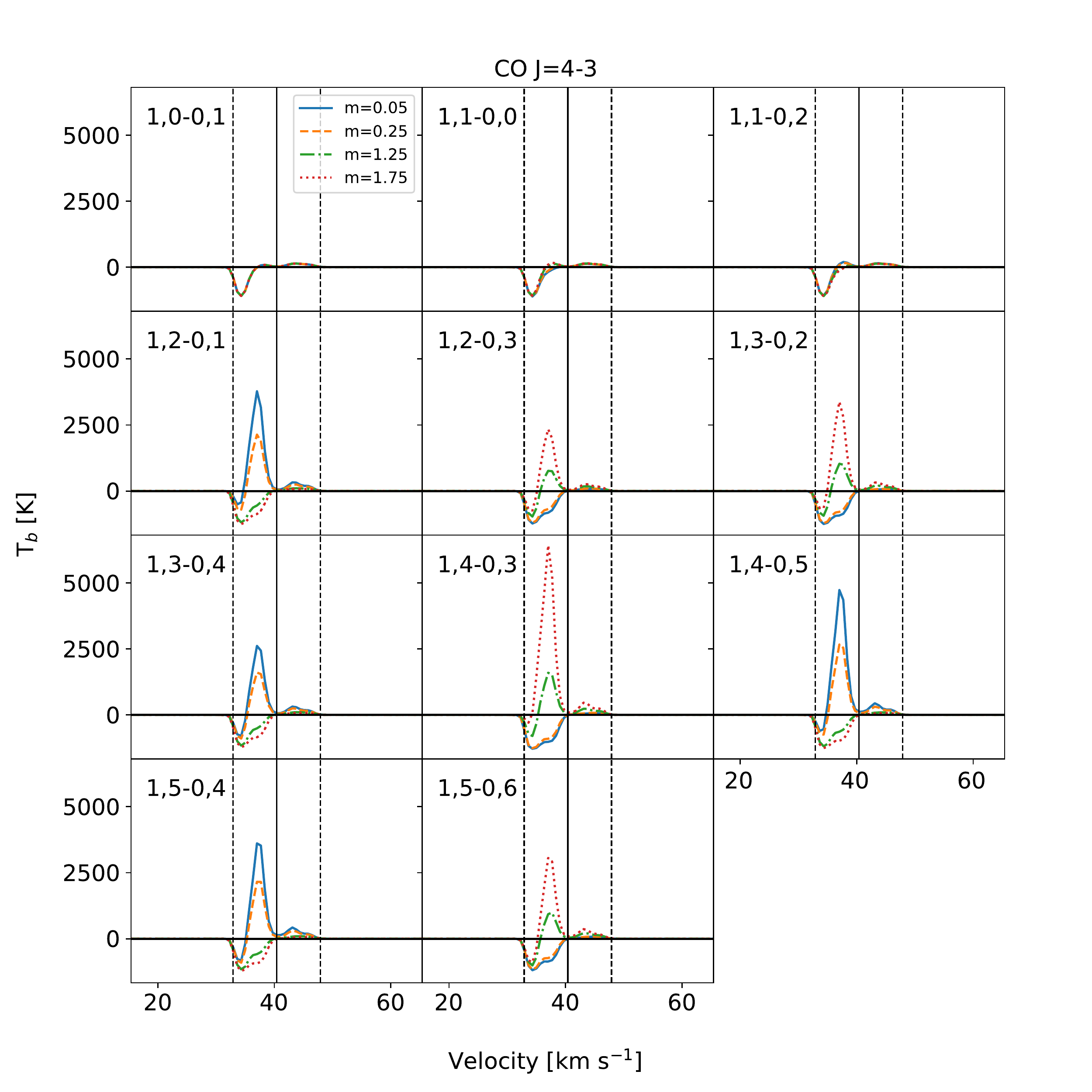}
\caption{As in Fig.~\ref{IRdiffspecs}, but without the atmospheric absorption component for the CO~$J=4-3$ emission observed towards the star with a 30~mas aperture. }
\label{IRdiffspecs43}
\end{figure*}

In Figs.~\ref{IRdiffspecs21},~\ref{IRdiffspecs32}, and \ref{IRdiffspecs43} we present the same models from \S~\ref{overlap}, including different relative radiation fields for the $v=0-1$ IR transitions, for the CO~$J=2-1$, $3-2,$ and $4-3$ lines. In these models no atmospheric absorption profile was included. 
\end{appendix}

\end{document}